# Correlation of the critical parameters of the gas-liquid phase transition and the density of the crystal at zero absolute temperature

*Umirzakov I.H.*

*Institute of Thermophysics, Novosibirsk, Russia*
*e-mail: tepliza@academ.org*

**Abstract**

The empirical Timmermann's formula connecting the density of the crystal at zero absolute temperature with the parameters of the critical point of the gas-liquid phase transition is derived theoretically from the Van der Waals equation of state using the conditions of phase equilibrium.

The equations of state $p(T,v) = kT/(v-b) - a/v^{\alpha}$ and $p(T,v) = kT/v + B/v^{\beta+1} - C/v^{\beta}$ with three parameters describing the critical point of any pure substance are considered. The Timmermann's formula is derived approximately from the first equation and exactly from the second one.

УДК 536.71, 536.22, 536.23

# Связь параметров критической точки фазового перехода газ - жидкость с плотностью кристалла при нулевой абсолютной температуре

© Умирзаков Ихтиёр Холмаматович

*Лаборатория моделирования, Институт теплофизики СО РАН,*
*проспект Лаврентьева, 1, 630090, г. Новосибирск, Россия*
*Тел.: (383) 354-20-17, E-mail:* [tepliza@academ.org](mailto:tepliza@academ.org)

**Ключевые слова**: *критическая температура, критический объем, критическое давление, критические параметры, уравнение состояния, фазовый переход жидкость-газ.*

## Аннотация

Эмпирическая формула Тиммерманса, связывающая плотность кристалла при нулевой абсолютной температуре с параметрами критической точки фазового перехода жидкость-газ, теоретически выведена на основе уравнения Ван-дер-Ваальса и условий фазового равновесия.

Рассмотрены трехпараметрические уравнения состояния $p(T,v) = kT/(v-b) - a/v^{\alpha}$ и $p(T,v) = kT/v + B/v^{\beta+1} - C/v^{\beta}$, позволяющие описать критическую точку любого однокомпонентного вещества. Формула Тиммерманса получена из первого уравнения состояния приближенно, а из второго уравнения - точно.

## Введение

Известно, что уравнение Ван-дер-Ваальса

$$p(T,v) = kT/(v-b) - a/v^2, \qquad (1)$$

где $p, T, v$ – давление, абсолютная температура и объем в расчете на одну молекулу (атом), $k$ – постоянная Больцмана, $a$ и $b$ – положительные параметры, является основополагающим для получения фундаментальных соотношений между термодинамическими параметрами, в частности, из него выводится уравнение идеальной кривой (линии единичной сжимаемости) [1]. В настоящей работе мы покажем, что из него можно вывести эмпирическую формулу Тиммерманса, связывающую параметры критической точки с плотностью кристалла при абсолютном нуле температуры.

## Вывод 1

Условия равенства давлений и температур сосуществующих в равновесии жидкости и пара дают уравнение

$$kT/(v_l(T)-b) - a/v_l^2(T) = kT/(v_g(T)-b) - a/v_g^2(T),$$

где $v_l(T)$, $v_g(T)$ - объем в расчете на одну частицу в жидкости и газе. Это уравнение можно привести к виду

$$kT/a \cdot 1/(1/v_l + 1/v_g) = 1 - b \cdot (1/v_l + 1/v_g) + b^2/v_l v_g. \qquad (2)$$

Учитывая, что в пределе низких температур в знаменателе дроби в левой части (2) можно положить $1/v_l + 1/v_g = 1/b$ (так как $v_l \to b$ и $v_g \to \infty$ при $T \to 0K$), а в правой части (2) можно пренебречь $b^2/v_l v_g \approx b/v_g$ по сравнению с $b/v_l$, из (2) получаем закон Кальете-Матиаса [2,с.292] (правило прямолинейного диаметра)

$$1/2 \cdot (1/v_l + 1/v_g) = 1/2b \cdot (1 - bkT/a). \qquad (3)$$

Этот закон применяется для определения значений плотности и температуры в критической точке, в частности он был успешно применен для определения критических параметров металлов [3]. Это правило является более общим, чем закон соответственных состояний [4]. Этот закон ранее был выведен другим способом в [5].

В критической точке уравнение состояния должно удовлетворять условию равенства давления его критическому значению

$$p(T_c, v_c) = p_c,$$

и условию [6,с.285]

$$[\partial p(T,v)/\partial v]_{v_c, T_c} = 0,$$

где $p_c$, $T_c$, $v_c$ - опытные значения давления, температуры и объема в критической точке. Наложив эти условия на уравнение состояния (1) легко получить уравнения

$$p_c v_c / kT_c = 1/(1 - b/v_c) - a/v_c kT_c,$$

$$a/v_c kT_c = 1/2(1 - b/v_c)^2, \qquad (4)$$

которые дают

$$p_c v_c / kT_c = (1 - 2b/v_c)/2(1 - b/v_c)^2. \qquad (5)$$

Из соотношения (3) имеем

$$a/v_c kT_c = b/v_c \cdot 1/(1 - 2b/v_c). \qquad (6)$$

Из (4) и (6) имеем

$$b/v_c = (1 - 2b/v_c)/2(1 - b/v_c)^2. \qquad (7)$$

Из сравнения (5) и (7) с учетом того, что $b$ равен объему $v_{00}$, занимаемому одним атомом (молекулой) в жидкости, переохлажденной до абсолютного нуля [7], имеем

$$p_c v_c / kT_c = v_{00} / v_c. \qquad (8)$$

Соотношение (8) известно как формула Тиммерманса [8,с.18]. Соотношение (8) позволяет определить $v_{00}$ через критические параметры, если его переписать в виде

$$v_{00} = p_c v_c^2 / kT_c.$$

Наиболее трудно определяемый критический параметр $v_c$ может быть получен с помощью формулы

$$v_c = \sqrt{v_{00} kT_c / p_c},$$

полученной из (8).

Формулу Тиммерманса (8) можно представить также как

$$p_c = kT_c v_{00} / v_c^2,$$

$$T_c = p_c v_c^2 / k v_{00},$$

$$p_c / T_c = v_{00} k / v_c^2.$$

Последние формулы позволяют определить критическое давление, критическую температуру и отношение критического давления к критической температуре.

## Вывод 2

Вводя новую переменную $\rho = 1/v$ уравнение Ван-дер-Ваальса (1) можно записать в виде

$$ab\rho^3 + a\rho^2 - (bp + kT)\rho + p = 0 \qquad (9)$$

Теорема Виета для корней $\rho_1 = \rho_L(T) = 1/v_L(T)$, $\rho_2$, $\rho_3 = \rho_G(T) = 1/v_G(T)$ кубического уравнения (9) относительно $\rho$ при $T \leq T_c = 8a/27kb$ даёт

$$\rho_1 + \rho_2 + \rho_3 = 1/b,$$

$$\rho_1 \rho_2 \rho_3 = p_e / ab,$$

где $p_e = \rho_e(T) = p(T, v_L) = p(T, v_G)$.

Из последних двух уравнений получаем

$$\rho_L + \rho_G = 1/b - p_e / \rho_L \rho_G ab \qquad (10)$$

С учетом (3) из (10) получаем

$$p_e v_L v_G = bkT \qquad (11)$$

При $T = T_c$ из соотношения (11) получаем формулу Тиммерманса (8).

## Вывод 3

Из (1) легко увидеть [1], что если $v = v_1(T)$, где $v_1(T)$ определяется из уравнения

$$bkT/a + b/v_1(T) = 1, \qquad (12)$$

то сжимаемость $pv/kT$ становиться равной единице. Поэтому линия определяемая уравнением (12) называется линией единичной сжимаемости. Кроме того эту линию также называют идеальной кривой, так как на этой кривой давление определяется из уравнения идеального газа.

В результате численного определения давления насыщенных паров, плотностей жидкости и газа Ван-дер-Ваальса из условий фазового равновесия нами получено приближенное соотношение

$$1/v_L + 1/v_G = 1/b - \alpha T. \qquad (13)$$

Уравнения (10) и (13) в критической точке дают

$$2/v_c = 1/b - p_c v_c^2 / ab,$$

$$2/v_c = 1/b - \alpha T_c,$$

Из двух последних уравнений имеем

$$\alpha = p_c v_c^2 / abT_c.$$

С учетом этого соотношения (13) принимает вид

$$1/v_L + 1/v_G = 1/b - (p_c v_c^2 / bkT_c) \cdot kT/a. \qquad (14)$$

При $T \leq 0.3 T_c$ для газа Ван-дер-Ваальса имеют место $|1/v_L - 1/v_G| < 10^{-3}/v_L$ и $|1/v_L - 1/v_1| < 10^{-2}/v_L$ поэтому из (12) и (14) имеем

$$1/v_1 = 1/b - kT/a,$$

$1/v_1 \approx 1/b - (p_c v_c^2 / bkT_c) \cdot kT/a$.

Последние два уравнения дают формулу Тиммерманса (8).

## Вывод 4

Рассмотрим обобщенное трехпараметрическое уравнение состояния Ван-дер-Ваальса-Дитеричи

$$p(T,v) = kT/(v-b) - A/v^{\alpha}, \qquad (15)$$

где $A$, $b$ и $\alpha$ - постоянные положительные параметры. В этом уравнении должно быть $\alpha > 1$ для того чтобы это уравнение состояния переходило в уравнение состояния идеального газа при малых плотностях. Оно при $\alpha = 5/3$ переходит в уравнение состояния Дитеричи, а при $\alpha = 2$ переходит в уравнение состояния Ван-дер-Ваальса ([9], стр.14-16).

Наложение на это уравнение условий равенства первой и второй частных производных давления по объему в расчете на одну частицу ($v$) в критической точке дает

$$v_c = b \frac{\alpha+1}{\alpha-1}, \qquad (16)$$

$$T_c = \frac{4 \cdot \alpha \cdot A \cdot b^{1-\alpha}}{k} \cdot \frac{(\alpha-1)^{\alpha-1}}{(\alpha+1)^{\alpha+1}}, \qquad (17)$$

которые с учетом (15) в критической точке дают

$$\frac{p_c v_c}{kT_c} = \frac{\alpha^2 - 1}{4\alpha}. \qquad (18)$$

Легко увидеть из (18), что критический фактор сжимаемости $z_c = \frac{p_c v_c}{kT_c}$ монотонно изменяется от нуля до бесконечности при изменении $\alpha$ от единицы до бесконечности, то есть уравнение состояния (15) способно описывать критическую точку любого однокомпонентного вещества.

Из уравнений (16) и (18) получаем

$$\frac{p_c v_c}{kT_c} = \frac{b/v_c}{1-(b/v_c)^2}.$$

Из него, получаем формулу Тиммерманса (8) при $(b/v_c)^2 \cong z_c^2 \ll 1$, выполняющегося с хорошей точностью для многих веществ.

Из (16)-(18) легко получить формулы, позволяющие определить параметры уравнения состояния (15) через критические параметры:

$$\alpha = 2z_c + \sqrt{1 + 4z_c^2},$$

$$b = v_c \cdot \frac{2z_c + \sqrt{1 + 4z_c^2} - 1}{2z_c + \sqrt{1 + 4z_c^2} + 1},$$

$$A = \frac{kT_c \cdot b^{\alpha-1}}{4 \cdot \alpha} \cdot \frac{(\alpha+1)^{\alpha+1}}{(\alpha-1)^{\alpha-1}}.$$

**Вывод 5**

Рассмотрим следующее уравнение состояния

$$p(T,v) = kT/v + B/v^{\beta+1} - C/v^{\beta}, \qquad (19)$$

где $B$, $C$ и $\beta$ - постоянные положительные параметры, причем должно быть $\beta > 1$, для того чтобы это уравнение состояния переходило в уравнение состояния идеального газа при малых плотностях. Оно может быть получено из уравнения состояния Путилова К.П. [10,11]

$$p(T,v) = kT/v + \overline{B}(T)/v^{n/3+1} - \overline{C}(T)/v^{m/3+1},$$

если в последнем положить $\overline{B}(T) = B = const$, $\overline{C}(T) = C = const$, $n = 3\beta$ и $m = 3\beta - 3$, соответственно.

Применяя условия равенства первой и второй частных производных давления по объему в расчете на одну частицу в критической точке к уравнению (19) получаем

$$v_c = \frac{B}{C}\frac{\beta+1}{\beta-1}, \qquad (20)$$

$$T_c = \frac{C}{kv_c^{\beta-1}}, \qquad (21)$$

$$z_c = \frac{\beta-1}{\beta+1}. \qquad (22)$$

Легко увидеть из (22), что критический фактор сжимаемости $z_c$ монотонно изменяется от нуля до единицы при изменении $\beta$ от единицы до бесконечности, то есть уравнение состояния (19) способно описывать критическую точку любого однокомпонентного вещества.

При абсолютном нуле температуры для холодного давления $p_0(v) = p(T=0,v)$ из (19) имеем

$$p_0(v) = B/v^{\beta+1} - C/v^{\beta},$$

из чего следует, что холодное давление обращается в нуль при

$$v_{00} = B/C. \qquad (23)$$

Из формул (20), (22) и (23) легко можно получить точную формулу Тиммерманса (8).

Формулы (20)-(22) позволяют легко получить формулы для определения параметров уравнения состояния (19) через критические параметры:

$$\beta = \frac{1+z_c}{1-z_c},$$

$$C = kT_c v_c^{\frac{1+z_c}{1-z_c}-1},$$

$$B = z_c kT_c v_c^{\frac{1+z_c}{1-z_c}}.$$

Отметим, что объем $v_{00}$ равен ортобарическому объему, то есть объему жидкости, находящейся в термодинамическом равновесии со своим паром, при экстраполяции к абсолютному нулю температуры [8,12]. Объем $v_{00}$ находится в тесной связи с «атомным объемом». Так, между $v_{00}$ и кубом атомного радиуса по Гольдшмидту имеет место прямая пропорциональность с погрешностью, составляющей в среднем порядка 2% [12]. $v_{00}$ равен объему, приходящемуся на один атом в идеальном кристалле при абсолютном нуле температуры, с точностью 1-2% [3,12]. В связи с этим нами выбрано такое название статьи.

## Заключение

Несмотря на то, что критическая точка была открыта более чем 140 лет назад, свойства вещества в этой точке окончательно не изучены и по сей день ведутся активные исследования в этом направлении (смотрите, например, [5,13-21]). При этом очень полезными оказываются соотношения, связывающие критические параметры с другими термодинамическими характеристиками, например с температурой Бойля [21].

В настоящей работе показано, что формула Тиммерманса, связывающая параметры критической точки с плотностью кристалла при абсолютном нуле температуры, может быть выведена из уравнения состояния Ван-дер-Ваальса.

Показано также, что рассмотренные нами в работе обобщенное уравнение состояния Ван-дер-Ваальса-Дитеричи и уравнение состояния Путилова приводят к формуле Тиммерманса.

Показано, что эти уравнения состояния могут описывать критическую точку любого однокомпонентного вещества.

Получены формулы, связывающие три критических параметра с тремя параметрами этих уравнений состояния.

Формула Тиммерманса получена пятью разными способами на основе трех различных уравнений состояния с использованием различных физических закономерностей: закона Кальете-Матиаса, уравнения линии единичной сжимаемости, условий фазового равновесия. Это позволяет утверждать, что эта формула неслучайная и имеет под собой веские физические основания.